\documentclass[a4paper,11pt]{article}

\pdfoutput=1 
\usepackage{jcappub}
\usepackage[T1]{fontenc}
\usepackage{comment}
\usepackage{multirow}
\usepackage{floatrow}
\usepackage{subfig}
\usepackage{graphicx}
\usepackage{url}

\title{\boldmath Generalized inflation in the context of $\kappa$-deformed theories}

\author[a,b,2]{B. W. Ribeiro,}
\author[a,1]{I. M. Macêdo,}
\author[a,3]{F. C. Carvalho.}

\affiliation[a]{Departamento de Física, Universidade do Estado do Rio Grande do Norte, Av. Professor Antonio Campos – Pres. Costa e Silva, Mossoró-RN, 59610-210, Brasil.}
\affiliation[b]{Departamento de Astronomia, Observatório Nacional, Rua General José Cristino 20921-400, Rio de Janeiro-RJ, Brasil.}

\emailAdd{brunoribeiro@on.br}
\emailAdd{isaacmacedo@alu.uern.br}
\emailAdd{fabiocabral@uern.br}

\abstract{A new inflationary scenario driven by a slowly-rolling homogeneous scalar field whose potential $V\left(\varphi\right)$ is 
given by a generalized exponential function is investigated.
Within the {\it slow-roll} approximation we obtain the main predictions of the model and compare them with current data from cosmic microwave background and large-scale structure observations. 
We show that this single scalar field model admits a wider set of solutions than usual exponential scenarios and predicts acceptable values of the spectral index, running of the spectral index and tensor-to-scalar ratio for the remaining number of {\it e}-folds lying in the interval $N = 55 \pm 5$ and an energy scale on which $\lambda \geq \sqrt{2}$; in particular, we observe that the value of the model parameter $\kappa$ depends on the analysis.
Finally, the primordial local non-Gaussianity is briefly discussed where we conclude that $k\gtrsim 0.02$ for $f_\text{NL}^\text{local} \ll 1$.}

\begin{document}
\maketitle
\flushbottom

\section{Introduction}
\label{sec1}
In the standard Hot Big Bang (HBB) cosmology, the cosmic inflation~\cite{Starobinsky_1980,Guth_1980,Linde_1983} is assumed to set the initial conditions driving the universe towards homogeneity, flatness, and absence of primordial monopoles, and generate the primordial inhomogeneities that lead to the formation of cosmic structures~\cite{Mukhanov_1981,Bardeen_1983}.
The cosmic inflation is a stage of accelerated expansion triggered by a homogeneous slowly-rolling scalar field $\varphi$, so-called {\it inflaton}, whose energy density is dominated by its potential $V(\varphi)$. Such an accelerated expansion took place in the very early universe and lasted just a split second, coming to the end once the kinetic and potential energy densities become comparable. 
Subsequently, the inflaton decays fully filling the universe with a hot plasma of relativistic particles, i.e, radiation, in a process known as reheating~\cite{Dolgov_1982,Abbott_1982,Nanopoulos_1983}.

During such an inflationary period, quantum fluctuations in the inflaton were stretched by ultrafast expansion and converted into cosmic microwave background (CMB) fluctuations, which are the seeds from which the large-scale structure (LSS) of the universe has evolved. Furthermore, the cosmic inflation predicts a stochastic background of primordial gravitational waves (PGWs)~\cite{Starobinsky_1979,Rubakov_1982,Abbott_1984} resulting from tensor perturbations in the metric, whose main observational signature is a {\it curl-like} pattern in the CMB polarization map, called B-modes~\cite{Baumann_2015,Kamionkowski_2016}. Since the universe before the last scattering is opaque to electromagnetic radiation, the PGWs represent a powerful window to probe the physics of the early universe and beyond.

In the framework of the simplest single field slow-roll inflation, both scalar and tensor perturbations are expected to have nearly scale-invariant spectra. A nearly scale-invariant spectrum is characterized by a specific pattern of fluctuations that are almost the same at all scales, i.e., they have compatible amplitudes and shapes regardless of the size of the region being observed. Such a spectrum can be completely described by a simple two-parameter power law. 
In addition, the initial spatial distributions of both scalar and tensor perturbations are nearly Gaussian, which means that all statistical information about them is contained in their two-point correlation functions or, equivalently, in their respective power spectra. These predictions have been confirmed by CMB (over the years by COBE~\cite{Smoot_1992}, WMAP~\cite{Bennett_2013}, and Planck~\cite{Akrami_2020_3} satellites) and LSS (2dFGRS~\cite{Colless_2001}, LSST~\cite{Abell_2009} and DES~\cite{Sayre_2020}) observations.

Since the temperature fluctuations measurements are close to the cosmic-variance limit for the tensor-to-scalar ratio, improvements are practically due to polarization. At present, a number of missions have been initiated to identify B-modes polarization in the CMB radiation and determine the tensor-to-scalar ratio. Among the most famous experiments, we highlight the ground-based 
AdvACT~\cite{Koopman_2018}, 
CLASS~\cite{Hileman_2014}, 
Keck/BICEP3~\cite{Grayson_2016}, 
Simons Array~\cite{Stebor_2016}, and 
SPT-3G~\cite{Benson_2014}, 
balloons-borne EBEX 10k~\cite{Abitbol_2018} and
SPIDER~\cite{Shaw_2020}, and 
satellites CMBPol~\cite{Smith_2009}, 
COrE~\cite{Caplan_2011} and
LiteBIRD~\cite{Hazumi_2020}.
Detecting the PGWs signal will lead the cosmology into a new era of discovery where it will be possible to go back to $\sim 10^{-32}\,$s after the Big Bang. As a result, it will be possible (albeit indirectly) to examine phenomena related to a new physics at the grand unification scale and the quantum behavior of the spacetime metric~\cite{Creminelli_2015}.

Despite the fact that the cosmic inflation is the cosmologists' favorite paradigm for explaining the history of the early universe and the formation of LSS, numerous questions remain unanswered. These include the following~\cite{Ijjas_2017}:
\begin{itemize}
    \item although many {\it toy models} of inflation have been proposed, embedding inflation into a realistic particle physics model has proven to be a challenge;

    \item primordial inflation requires that the universe be dominated by the energy density of a purely hypothetical ingredient, referred to as inflaton, for which there is no direct evidence;

     \item if inflation is generated by a scalar matter field, a spacetime singularity must have preceded this period meaning that the inflationary cosmology is incomplete;
    
    \item the theory of cosmological fluctuations is based on the General Relativity theory, that is without doubt inapplicable at the sub-Planckian scales present during the early phases of inflation; and

    \item have not been found (to date) any signs of the PGWs background, predicted by the cosmic inflation, despite meticulous searches for them.
\end{itemize}
The above-mentioned conceptual difficulties are the main motivations for searching for some alternative theories to primordial inflation. The most recurrent in the literature are: 
varying light speed~\cite{Albrecht_1999}, 
topological defects~\cite{Vilenkin_2000}, 
ekpyrotic and cyclic cosmology~\cite{Lehners_2008}, 
string gas~\cite{Battefeld_2006}, and 
bouncing cosmologies~\cite{Brandenberger_2017}. 
As we can see, none of these alternatives theories is as well developed as inflationary cosmology at present, and none of them solves the standard HBB cosmology problems as elegantly as cosmic inflation does. Furthermore, as previously stated, the most recent and robust CMB and LSS observations are in fully agreement with the inflationary predictions.

It should be emphasized that the detailed particle physics mechanism responsible for the primordial inflation is still unknown, so any description of this era requires an appropriate extrapolation of the known physical laws. In consequence, the standard approach is strictly phenomenological with a considerable freedom in modeling the inflationary potential. In this sense, several shapes of the potential $V(\varphi)$ have been proposed in the literature giving rise to a number of inflationary models, from conservative approaches like power law to alternative approaches like tachyon and ghost (see refs.~\cite{Bassett2006,Baumann_2009,Lyth_2009} for a review on inflation).

A simple and interesting class of inflationary models is that described through the usual exponential potential
\begin{equation}\label{inflacaoexp}
V\left(\varphi\right)\propto \exp\left(-\lambda\varphi\right) \,,
\end{equation}
as originally investigated in refs.~\cite{Peebles_1988,Ratra_1988}.
It turns out that scalar fields with exponential potentials are very common in a number of particle physics theories. Indeed, they naturally occur in supergravity~\cite{Salam_1984}, 
superstrings~\cite{Easther_1993}, 
higer-order~\cite{Wetterich_1989}, 
higher-dimensional~\cite{Shafi_1984}, and 
Kaluza-Klein theories~\cite{Wehus_2004}, 
among others. 
In this paper we propose a generalization of the usual exponential scenario in eq.~(\ref{inflacaoexp}) in light of the $\kappa$-deformed functions~\cite{Kaniadakis_2002}. 
Generalized cosmological scenarios of early and late universe have been previously investigated in ref.~\cite{Carvalho_2006}. In particular, it has been shown that the $\beta$-exponential potential proposed in ref.~\cite{Alcaniz_2007} can be derived from the framework of the braneworld scenarios and that it provides a good description of the observational data~\cite{dos_Santos_2022}. On the other hand, the $\kappa$-generalized scenario studied in ref.~\cite{Lambiase_2023} has been proposed in the context of the gravity-thermodynamics conjecture~\cite{Padmanabhan_2010}, such that a new cosmological scenario emerges based on modified Friedmann equations. In contrast, the $\kappa$-generalization of the ordinary exponential inflationary scenarios that we are proposing here occurs within the standard hypothesis, i.e., without modifying the usual Friedmann equations or introducing new thermodynamic conjectures concerning the spacetime structure.

This paper is organized as follow. In section~\ref{din.inf.} we introduce the formal basis of the inflationary dynamics in the framework of the slowly-rolling homogeneous scalar field. In section~\ref{cosm.pert.} we consider quantum fluctuations in the inflaton and develop mathematical tools in order to characterize them statistically, such as the primordial power spectrum, spectral index, running of the spectral index, and tensor-to-scalar ratio. In addition, we also provide a brief description of possible primordial non-Gaussianities (NGs) through the derivation of the non-linearity parameter. In section~\ref{k-exp.inf.} we present our $\kappa$-generalized model, its main predictions and compare our results with the current Planck CMB and LSS data. Finally, we discuss our results in light of the current status of the cosmic inflation in section~\ref{dic.}.

\section{\label{din.inf.}Dynamics of Inflation}
In this section we shall develop the main concepts regarding the physics of the scalar field, which is assumed to be the main driver of the primordial inflation. In this way we will show how such a scalar field produces an almost de~Sitter expansion, in early times, required to explain the current features of the universe. Within the slow-roll approximation, we shall define some of the main inflationary parameters and discuss general aspects of this approach.

\subsection{Physics of the scalar field}
The simplest dynamical model of inflation involves a homogeneous single scalar field $\varphi(t)$, called inflaton, minimally coupled with the gravity and characterized by the potential $V(\varphi)$. Such a theory is described by the action~\cite{Baumann_2009} 
\begin{equation}\label{action}
    S = \int{d^4 x\sqrt{-g}}\left[ \frac{1}{2}R + \frac{1}{2}g^{\mu\nu}\partial_\mu\varphi\partial_\nu\varphi - V(\varphi) \right]\,,
\end{equation}
in units for which $M_{\rm Pl}^2 \equiv (8\pi G)^{-1} = \hbar = c = 1$.

In the present context, the Friedmann-Lemaître-Roberton-Walker (FLRW) metric ansatz fits the global properties about the statistical homogeneity and isotropy of the universe, i.e.,
\begin{equation}\label{metrica}
{ds}^2= - {dt}^2 + a^2(t)\left(\frac{{dr}^2}{1-Kr^2} + r^2{d\Omega}^2\right), \quad K = 0, \pm 1,
\end{equation}
where $K$ is the spatial curvature parameter and $a(t)$ is the scale factor. Assuming a spatially flat ($K=0$) 3-section and a perfect fluid treatment, the stress-energy tensor corresponding to the extra scalar field $\varphi$ is described by
\begin{equation}
    T_{\mu\nu}^{(\varphi)} = - \frac{2}{\sqrt{-g}}\frac{\delta S_\varphi}{\delta g^{\mu\nu}} = \partial_\mu\varphi\partial_\nu\varphi - g_{\mu\nu}\left[\frac{1}{2}\partial_\rho\varphi\partial^\rho\varphi + V(\varphi)  \right]\,.
\end{equation}
For a homogeneous field configuration ($\nabla \varphi$ = 0), this leads to the energy density and pressure
\begin{equation}
    \rho_\varphi = \frac{1}{2}\dot{\varphi}^2 + V(\varphi)\,,
\end{equation}
\begin{equation}
    P_\varphi = \frac{1}{2}\dot{\varphi}^2 - V(\varphi)\,,
\end{equation}
respectively, where overdots indicate derivative with respect to the cosmic time $t$. Hence, the equation of state (EoS) corresponds to
\begin{equation}
    w_\varphi=\frac{P_\varphi}{\rho_\varphi} = \frac{\frac{1}{2}\dot{\varphi}^2 - V(\varphi)}{\frac{1}{2}\dot{\varphi}^2 + V(\varphi)}\,.
\end{equation}

Then varying the action in eq.~(\ref{action}) with respect to the scalar field $\varphi$, it is possible to arrive at the Klein-Gordon equation,
\begin{equation}\label{klein-gordon}
    \ddot{\varphi} + 3H\dot{\varphi} + V^{\prime}(\varphi) = 0\,,
\end{equation}
where primes indicate the derivative with respect to the field $\varphi$. On the other hand, varying this action with respect to the metric tensor $g_{\mu\nu}$, we get the Friedmann equations,
\begin{equation}\label{friedmann1}
    H^2 = \frac{1}{3}\left[\frac{1}{2}\dot{\varphi}^2 + V(\varphi)\right]\,,   
\end{equation}
\begin{equation}\label{friedmann2}
    \frac{\ddot{a}}{a} = -\frac{1}{3}\left[\dot{\varphi}^2 - V(\varphi)\right]\,.
\end{equation}
Eqs.~(\ref{friedmann1}) and~(\ref{friedmann2}) govern the evolution of the universe filled by a homogeneous scalar field $\varphi(t)$ whose dynamics is determined by eq.~(\ref{klein-gordon}). In this way, a complete and suitable solution involves solving all three equations at the same time.

\subsection{Slow-roll approximation}
Notice then that a dynamical homogeneous scalar field $\varphi(t)$ may induce an early inflationary epoch -- without undermining the successes of the standard HBB cosmology -- provided that the potential term $V(\varphi)$ is dominant over the kinetic term, i.e., $\dot{\varphi}^2 \ll V$, and sufficiently flat, i.e.,  $V^\prime, V^{\prime\prime} \ll V$. In addition, inflation will only be sustained for a sufficiently long period of time if the second time derivative of $\varphi$ is small enough, i.e., $|\ddot{\varphi}| \ll 3H\dot{\varphi}$. This conditions are known as slow-roll conditions, leading to the approximation~\cite{Baumann_2009}
\begin{equation}
    V^{\prime} \simeq -3H\dot{\varphi}\,,
\end{equation}
\begin{equation}
    H^2 \simeq \frac{1}{3}V(\varphi)\,,
\end{equation}
\begin{equation}
    \dot{H} \simeq - \frac{1}{2}\dot{\varphi}^2\,.
\end{equation}
In this case, the dynamics of inflation can be expressed in terms of the slow-roll parameters as defined by~\cite{Lyth_2009}
\begin{eqnarray}
    \epsilon &=& \frac{1}{2}\left(\frac{V^{\prime}}{V}\right)^2 \,, \label{epv}\\
    \eta &=& \frac{V^{\prime\prime}}{V} - \frac{1}{2}\left(\frac{V^{\prime}}{V}\right)^2 \,, \label{etav}\\
    \xi^2 &=& \frac{V^{\prime} V^{\prime\prime\prime}}{V^2} \label{xiv}\,. 
\end{eqnarray}
In order to eqs.~(\ref{klein-gordon})--(\ref{friedmann2}) are in agreement with the slow-roll approximation, the above slow-roll parameters must be extremely small, i.e., 
\begin{equation}
    \epsilon \ll 1\,, \quad |\eta| \ll 1\,, \quad \xi^2 \ll 1\,. 
\end{equation}
However, an accelerated expansion phase requires only that $\ddot{a} > 0$ implying $\epsilon < 1$. Hence, an inflationary period ends when $\epsilon = 1$.

When second-order contributions in the primordial power spectra are included, Planck TT, TE, EE+lowE+lensing(+BK15) data constrain the slow-roll parameters\footnote{Planck collaboration uses a definition of the slow-roll parameters which relate to ours via the relations $\epsilon_V = \epsilon$, $\eta_V=\epsilon_V + \eta$, and $\xi^2_V=\xi^2$. Thus, the subscript $V$ in eq.~(\ref{Planck_sr_par}) refers to the slow-roll parameters as set and estimated by Planck.} as~\cite{Akrami_2020_3}
\begin{equation}\label{Planck_sr_par}
    \begin{array}{llll}
    \epsilon_V < & 0.0097 &\quad (0.0044) &\quad \text{at} \ 95\,\%~\text{CL}\,,\\
    \\[-3.25mm]
    \eta_V = & -0.010^{+0.007}_{-0.011} &\quad (-0.015 \pm 0.006) &\quad \text{at} \ 68\,\%~\text{CL}\,, \\
    \\[-3.25mm]
    \xi^2_V =& 0.0035^{+0.0078}_{-0.0072} &\quad (0.0029^{+0.0073}_{-0.0069}) &\quad \text{at} \ 95\,\%~\text{CL}\,.
    \end{array}
\end{equation}
As we will see later, these constraints -- especially those on $\epsilon$ -- combined to the tensor-to-scalar ratio and the amplitude of the scalar fluctuations, provide an upper bound on the energy scale of inflation when the pivot scale $k_*$ exits the Hubble radius.

\subsection{Number of {\it e}-folds}
Within the slow-roll approach, it is simple to calculate the scale factor at any instant between the beginning and the end of inflation. Since the expansion is assumed to be very large in this stage, it is more convenient to compute it in terms of the number of {\it e}-folds remaining before the end of inflation, defined as~\cite{Baumann_2009} 
\begin{equation}\label{e-folds}
    N \equiv \ln{\left(\frac{a_\text{end}}{a}\right)} = \int_{\varphi_\text{end}}^{\varphi}{\frac{d\varphi}{\sqrt{2\epsilon(\varphi)}}}\,,
\end{equation}
where $\varphi_\text{end}$ denotes the value of the scalar field at the end of inflation. This quantity measures then the amount of physical expansion during inflation.

To determine the number of {\it e}-folds corresponding to a particular scale $k$ in terms of the present Hubble scale $k_0 =a_0H_0$, we need a model describing the complete history of the Universe. In the standard HBB scenario, inflation is followed by a period of reheating, then a period dominated by radiation, then one dominated by non-relativistic matter, and finally the current one dominated by the cosmological constant. In this background, assuming instantaneous transitions between one era to another, then one has~\cite{Akrami_2020_3}
\begin{equation}
    N(k) \simeq 67 - \ln{\left(\frac{k}{a_0H_0}\right)} + \frac{1}{4}\ln{\left(\frac{V_k^2}{\rho_\text{end}}\right)} - \frac{1-3w_\text{int}}{12(1 + w_\text{int})}\ln{\left(\frac{{\rho_\text{end}}}{\rho_\text{th}}\right)} - \frac{1}{12}\ln{\left(g_\text{th}\right)}\,,
\end{equation}
where $V_k$ is the potential energy when the scale $k$ crosses the Hubble radius during inflation ($k = a_kH_k$), $\rho_\text{end}$ is the energy density at the end of inflation, $w_\text{int}$ corresponds to the effective EoS between the end of inflation and the thermalization energy scale $\rho_\text{th}$, and $g_\text{th}$ is the number of effective bosonic degrees of freedom at the energy scale $\rho_\text{th}$. Planck collaboration assumes $g_\text{th}=10^3$, a  pivot scale $k_*=0.002~\text{Mpc}^{-1}$, and an uncertainty of $50 < N_* < 60$~\cite{Akrami_2020_3}.

\section{\label{cosm.pert.}Cosmological Perturbations}
Although inflation arose in a context in which the main problems to be solved were those of flatness, horizon and cosmic relics, the most robust inflationary prediction is the generation of density fluctuations as seeds for LSS in the Universe. Quantum fluctuations of the inflaton field are stretched on large scales by the accelerated expansion and frozen after the scale of perturbations leaves the Hubble radius during inflation. Long after inflation ends, the perturbations cross the Hubble radius again providing a natural explanation for the observed anisotropies in the CMB and the LSS observed today~\cite{Starobinsky1982,Stewart1993}. On the other hand, if the energy scale of inflation is high enough, this mechanism is also expected to generate a subtle background of PGWs that can polarize the CMB photons, leading to a very distinctive signature in the B-modes spectrum on large angular scales.~\cite{Mukhanov2005, Bassett2006}.

Fundamentally, we are interesting in only small fluctuations around the homogeneous single scalar field $ \varphi(t, x^i) = \bar{\varphi}(t) + \delta\varphi(t, x^i)$, where from now on we will use a bar in order to identify background quantities and neglect non-linear terms in the perturbations. 
The inflationary field has been assumed to be minimally coupled to the gravity. This means that any perturbation $\delta\varphi$ will induce perturbations on the metric $\delta g_{\mu\nu}$ which are divided into scalar, vector and tensor modes according to the scalar-vector-tensor (SVT) decomposition. In this way, metric perturbations sourced by the fluctuations of the inflaton can be described by the perturbed FLRW line element~\cite{Bassett2006},
\begin{equation}
    {ds}^2 = - (1+2\Phi){dt}^2 + 2aB_idtdx^i + a^2[(1-2\Psi)\delta_{ij}+E_{ij}]dx^idx^j\,,
\end{equation}
where $B_i = \partial_iB-S_i$ and $E_{ij} = 2\partial_{ij}E + 2\partial(_iF_j) + h_{ij}$. 
Since the perturbations are decoupled at the linearized level, they evolve separately~\cite{Bardeen1980,Mukhanov1992,Kodama1984}. 
Scalar fluctuations are often described in terms of the comoving curvature perturbation $\mathcal{R} = \Psi + (H/\dot{\bar{\varphi}})\delta\varphi$, obeying the following equation,
\begin{equation}\label{escalarmodes}
    \frac{1}{a^3\epsilon}\frac{\partial}{\partial t}\left(a^3\epsilon\dot{\mathcal{R}}\right) + \frac{k^2}{a^2}\mathcal{R} = 0\,.
\end{equation}
Tensor fluctuations in turn can be described in terms of the decomposition $h_{ij}=h(t)e_{ij}^{(+,\times)}$, where $e_{ij}$ are the eigenmodes of the spatial Laplacian operator, and $h(t)$ is some amplitude term obeying the following equation,
\begin{equation}\label{tensormodes}
    \ddot{h} + 3H\dot{h} + \frac{k^2}{a^2}h = 0\,.
\end{equation}
Thus, tensor fluctuations give rise to PGWs with two possible polarization states, $+$ and $\times$, and time-dependent amplitude $h=h(t)$. Vector fluctuations are expected to vanish in the presence of only scalar fields~\cite{Bassett2006}.

Since the primordial scalar and tensor perturbations are expected to be nearly Gaussian, they can be described in terms of their two-point correlation functions, 
\begin{equation}    
\left<\mathcal{R}_k\,\mathcal{R}_{k^\prime}\right> = (2\pi)^3\delta^3\left(k+k^\prime\right) P_{\rm s}(k)\,, 
\end{equation}
\begin{equation}
    \left<h_k\,h_{k^\prime}\right> =  (2\pi)^3\delta^3\left(k+k^\prime\right) P_{\rm T}(k)\,,
\end{equation}
where $P_{\rm s}(k)$ and $P_{\rm T}(k)$ are the scalar and tensor power spectra, respectively. Nevertheless, throughout this paper we will work in terms of the dimensionless spectra\footnote{The dimensional and dimensionless power spectra are connected via the relation $\mathcal{P}_i(k) = \frac{k^3}{2\pi^2}P_i(k)$, where $i$ denotes both scalar (s) and tensor (T) perturbations.} in the framework of the single field slow-roll approximation, which are respectively given by
\begin{equation}\label{scalarspectrum}
    \mathcal{P}_{\rm s}(k) = \frac{4\pi k^3}{(2\pi)^3}\left|\mathcal{R}^2\right|
   =\left(\frac{1}{12\pi^2}\right)\left(\frac{V^3}{{V^\prime}^2}\right)\,,
\end{equation}
\begin{equation}\label{tensorspectrum}
    \mathcal{P}_{\rm T}(k) = 2\frac{4\pi k^3}{(2\pi)^3}\left|h^2\right|
    = \left(\frac{1}{12\pi^2}\right) \left(\frac{V^3}{{V^\prime}^2}\right)\,,
\end{equation}
where the factor $2$ in eq~(\ref{tensorspectrum}) encodes the two independent polarizations of the graviton, and $\mathcal{R}$ and $h$ are the solutions of eqs.~(\ref{escalarmodes}) and~(\ref{tensormodes}) respectively. Here, both power spectra are evaluated at the Hubble-radius crossing, $k=aH$. Since both scalar and tensor perturbations are sourced by the inflaton quantum fluctuations in an almost de Sitter background, we must expect the primordial power spectra to be nearly flat.

For single field models, the scale-dependence of the primordial power spectra can be approximated by a power-law form of the adiabatic scalar and tensor components, thus the scalar and tensor spectral indices are  
\begin{equation}
    n_{\rm s} - 1 = \left[ \frac{d\ln{\mathcal{P}_{\rm s}(k)}}{d\ln{k}}  \right]_{k=k_*}
\end{equation}
and
\begin{equation}
    n_{\rm T} = \left[ \frac{d\ln{\mathcal{P}_{\rm T}(k)}}{d\ln{k}}  \right]_{k=k_*}\,,
\end{equation}
respectively. Here, $k_*$ denotes an arbitrary pivot scale. Notice that the scale-invariant power spectra correspond to $n_{\rm s}=1$ and $n_{\rm T}=0$~\cite{Martin2023}.

Perturbations generated in a single field slow-roll regime are expected to be only weakly scale-dependent. Using eqs.~(\ref{scalarspectrum}) and (\ref{tensorspectrum}) we can write the scalar and tensor spectral indices in terms of the  slow-roll parameters as
\begin{equation}\label{nsh}
    n_{\rm s} - 1 = 2\eta - 4\epsilon
\end{equation}
and
\begin{equation}
    n_{\rm T} = -2\epsilon\,.
\end{equation}
Indeed, Planck TT, TE, and EE+lowE+lensing data give us $n_{\rm s} = 0.9649 \pm 0.0042$ at $68\,\%$ CL in the base-$\Lambda$CDM model~\cite{Akrami_2020_3}, confirming a nearly flat power spectrum just as expected in the single field slow-roll inflation. This scenario do not significantly changes when we consider $\Lambda$CDM extensions. For instance, for the base-$\Lambda\text{CDM}+\alpha_{\rm s}+\beta_{\rm s}$ model, Planck TT, TE, EE+lowE+lensing data give us $n_{\rm s} = 0.9625 \pm 0.0048$ at $68\,\%$ CL~\cite{Akrami_2020_3}. Direct detection of tensor modes has not yet been achieved.

\subsection{Running of the spectral index}
Because there is a weak scale-dependence on the spectral indices of both scalar and tensor perturbations, we shall seek a way to measure them. Considering the third terms proportional to $(1/2)\ln^2{(k/k_*)}$ in the power spectra expansions\footnote{It turns out that the power spectra of the primordial fluctuations can be Taylor-expanded in terms of $\ln{k}$ around the pivot scale $k_*$ as $\ln{\mathcal{P}_i(k)} = \ln{\mathcal{P}_i(k_*)} + \frac{d\ln{\mathcal{P}_i(k_*)}}{d\ln{k}}\ln{\left(\frac{k}{k_*}\right)} + \frac{1}{2}\frac{d^2\ln{\mathcal{P}_i}(k_*)}{d\ln{k}^2}\ln^2{\left(\frac{k}{k_*}\right)} + \cdot\cdot\cdot$~\cite{Zarei2016}.}, the respective coefficients are called running of the scalar and tensor spectral indices, defined as~\cite{Zarei2016}
\begin{equation}\label{running}
    \alpha_{\rm s} = \left[\frac{d n_{\rm s}}{d\ln{k}}\right]_{k=k_*}\,, \quad   \alpha_{\rm T} = \left[\frac{d n_{\rm T}}{d\ln{k}}\right]_{k=k_*}\,.
\end{equation}
Since $(d/d\ln{k}) = (1/H)(d/dt)$, we can say that the running of the spectral index quantify the rate of change of $\{n_{\rm s},n_{\rm T}\}$ per Hubble time. In terms of the slow-roll parameters we have~\cite{Bassett2006}  
\begin{equation}
    \alpha_{\rm s} = 16\epsilon\eta - 8\epsilon^2 - 2\xi^2\,,
\end{equation}\label{running2}
\begin{equation}
    \alpha_{\rm T} = -4\epsilon\left(\epsilon - \eta\right)\,.
\end{equation}
Notice then that, in this scenario, $\alpha_{\rm s},\alpha_{\rm T} \ll 1$. Indeed, Planck 2018 data are consistent with the single field slow-roll inflation as it has constrained $\alpha_{\rm s} = 0.002 \pm 0.010$ at $68~\%$ CL in the base-$\Lambda\text{CDM}+\alpha_{\rm s}+\beta_{\rm s}$ model~\cite{Akrami_2020_3}.

\subsection{Tensor-to-scalar ratio}
An important observational quantity is the called tensor-to-scalar ratio $r$, defined as the ratio between amplitudes of the tensor and scalar power spectra at pivot scale $k_*$, i.e.,
\begin{equation}
    r = \frac{\mathcal{P}_{\rm T}(k_*)}{\mathcal{P}_{\rm s}(k_*)} = \frac{A_{\rm T}}{A_{\rm s}}\,.
\end{equation}
In the single field slow-roll framework, we obtain
\begin{equation}\label{ratio2}
    r = 16\epsilon = - 8 n_{\rm T}\,,
\end{equation}
where the second equality is called consistence relation, being a single field slow-roll inflation prediction~\cite{Lidsey1997}. Since $\epsilon \ll 1$ in the slow-roll regime, we expect that the amplitude of the tensor modes to be suppressed by the scalar modes. However, it must be possible to observe them. In a universe dominated by the energy density of a single slowly-rolling scalar field, the total field excursion is related to the tensor amplitude by $\Delta\varphi/M_{\rm Pl} = \sqrt{r/8}\,N$~\cite{Lyth1997}. Thus, for $N=50$ we get the lower bound~\cite{Forconi2021}
\begin{equation}
    \frac{\Delta\varphi}{M_{\rm Pl}} = \sqrt{\frac{r}{3.2 \times 10^{-3}}}.
\end{equation}
Notice then that a large value for the tensor-to-scalar ratio, $r\geq 3.2 \times 10^{-3}$, is associated with both significant amplitude of the tensor perturbations and a high scale for the inflationary energy, $\Delta\varphi \geq M_{\rm Pl}$. Therefore, single field slow-roll inflation generates a PGWs background which may be observable in the low multipoles of the CMB anisotropies, but only if the scalar field variation is at least of the order of the Planck scale~\cite{Lyth1997}.

Without observing the tensor modes, Planck was only able to infer the following upper limits considering the base-$\Lambda\text{CDM}+r$ model~\cite{Akrami_2020_3}: 
\begin{itemize}
    \item $r<0.11$ and $r_{0.002}<0.10$ for Planck TT, TE, EE+lowEB+lensing; and
    \item $r<0.061$ and $r_{0.002}<0.056$ for Planck TT, TE, EE+lowE+lensing+BK15,
\end{itemize}
On the other hand, the following restrictions have been obtained by Forconi {\it et al.}~\cite{Forconi2021} considering the base-$\Lambda\text{CDM}+r+\alpha_{\rm s}$ model:
\begin{itemize}
    \item $r<0.159$ for Planck TT, TE, EE+lowEB+lensing; and
    \item $r<0.0658$ for Planck TT, TE, EE+lowE+lensing+BK15.
\end{itemize}
In any case, the constraints on this parameter are at $68\,\%$~CL, while the upper bounds are at $95\,\%$~CL. Therefore, the Planck 2018 baseline plus BK15 constraint on $r$ implies in the upper bound on the energy scale of inflation~\cite{Akrami_2020_3},
\begin{equation}
    V_* = \frac{3\pi^2M_\text{Pl}^4}{2} A_{\rm s} r < \left( 1.6 \times 10^{16}\,\text{GeV} \right)^4 \quad (95\,\%~\text{CL})\,, 
\end{equation}
or, equivalently, 
\begin{equation}
    \frac{H_*}{M_\text{Pl}} < 2.5 \times 10^{-5} \quad (95\,\%~\text{CL})\,.
\end{equation}
Additionally, Forconi {\it et al.}~\cite{Forconi2021} have obtained constraints on the PGWs considering non-vanishing curvature, $\Omega_{\rm K} \neq 0$, and ACTPol,  SPT-3G and WMAP data combinations. Anyway, the results do not significantly change from those explained above.

\subsection{Primordial non-Gaussianity}
Curvature perturbations generated by quantum fluctuations in an inflationary phase produce a quasi-Gaussian random density field. Nevertheless, detectable amounts of non-Gaussianity (NG) can be produced in scenarios such as inflaton self-interactions, additional light and/or heavy fields, and multi-fields. Thus even a small amount of primordial NG would be a sufficient argument to go beyond the simplest single field slow-roll scenario. In this way, primordial NG ends up being an important tool for discriminating between inflation theories~\cite{Meerburg2019,Floss2023}.

Non-Gaussianity is statistically characterized by a non-vanish three-correlation point function, or its Fourier transform, the bispectrum $B_{\rm s}$, defined as~\cite{Baumann_2009}
\begin{equation}
    \left< \mathcal{R}_{k_1}\mathcal{R}_{k_2}\mathcal{R}_{k_3} \right> = \left(2\pi\right)^3\delta^3\left(k_1 + k_2 + k_3\right) B_{\rm s}\left(k_1, k_2, k_3\right)\,.
\end{equation}
There are three main bispectrum templates most explored in the literature, the so-called local, equilateral and orthogonal shapes. In this work we shall explore the local template.

The simplest way in order to parameterize possible primordial NGs is through non-linear corrections to the Gaussian curvature perturbation $\mathcal{R}_{\rm G}$, i.e.,
\begin{equation}
    \mathcal{R} = \mathcal{R}_{\rm G} + \frac{3}{5}f_\text{NL}^\text{local}\left[\mathcal{R}_{\rm G}^2 - \left<\mathcal{R}_{\rm G}^2\right>\right]\,,
\end{equation}
where $f_\text{NL}^\text{local}$ quantify the degree of local non-linearity on the perturbation $\mathcal{R}$\footnote{The curvature perturbation $\mathcal{R}$ is related to the primordial potential $\Phi$ in the matter-dominated epoch as  $\Phi = (3/5)\mathcal{R}$~\cite{Baumann_2009}.}. In this way, it is easy to show that
\begin{equation}
    f_\text{NL}^\text{local} = \frac{5}{6} \left[ \frac{B_{\rm s}\left(k_1, k_2, k_3\right)}{\mathcal{P}_{\rm s}\left(k_1\right)\mathcal{P}_{\rm s}\left(k_2\right) + \mathcal{P}_{\rm s}\left(k_2\right)\mathcal{P}_{\rm s}\left(k_3\right) + \mathcal{P}_{\rm s}\left(k_3\right)\mathcal{P}_{\rm s}\left(k_1\right)} \right]\,.
\end{equation}
In the single field slow-roll approach, both inflationary power spectrum and bispectrum can be obtained from the $\delta N$ formalism~\cite{Sasaki1996,Sugiyama2013}, so that the non-linearity parameter gives~\cite{Bassett2006,Wands2010}
\begin{equation}\label{fNLH}
    f_\text{NL}^\text{local} = - \frac{5}{6} \left(\eta - 2\epsilon\right) = - \frac{5}{12} \left(n_{\rm s} - 1\right)\,.
\end{equation}
The second equality is the second consistence relation for the single field slow-roll inflation. As we can see, the local NG is completely suppressed by the slow-roll parameters, or equivalently $n_{\rm s} - 1$, in the slow-roll inflation. Therefore, any robust measurement revealing a large $f_\text{NL}^\text{local}$ rules out the single field inflation~\cite{Maldacena2003,Paolo2004}. 

\section{\label{k-exp.inf.}\texorpdfstring{$\kappa$}{kapa}-Exponential Inflation}
In this section we propose a new inflationary model, the $\kappa$-exponential inflation. This model generalizes the conventional exponential scenarios originally studied in refs.~\cite{Peebles_1988,Ratra_1988}, described in eq.~(\ref{inflacaoexp}). In the scope of the slow-roll approximation, we derive the main predictions of the model, such as the slow-roll parameters, number of {\it e}-folds, scalar spectral index and its running, tension-to-scalar ratio, and primordial NGs, and its cosmological consequences. Finally, we confront our results with the current observations.

\subsection{The model}
In the context of the $\kappa$-deformed theories, we propose the following generalization of the inflationary potential in eq.~(\ref{inflacaoexp}), given by
\begin{equation}\label{kpot}
V(\varphi) \propto \exp_{\kappa}(-\lambda\varphi) \,,
\end{equation}
where the above generalized exponential and its inverse, called $\kappa$-logarithmic, are defined as
\begin{equation}
    \exp_{\kappa}(x) \equiv (\sqrt{1+\kappa^2 x^2} + \kappa x)^{\frac{1}{\kappa}}\,,
\end{equation}
\begin{equation}
    \ln_{\kappa}{(x)} \equiv \frac{x^{\kappa} - x^{-\kappa}}{2\kappa}\,,
\end{equation}
which reduce to ordinary exponential and logarithm functions, respectively, as the parameter $\kappa \in [-1,1]$ approaches zero. Notice, $x \equiv -\lambda\varphi$ for our purposes right here~\cite{Kaniadakis_2013,Kaniadakis_2001,Kaniadakis_2002}.

Some properties of the ordinary exponential and logarithmic functions are preserved in the above generalized approach. For instance, we have
\begin{equation}
    \exp_{\kappa}\left[\ln_{\kappa}{(x)}\right] = x\,,
\end{equation}
\begin{equation}
    \exp_{\kappa}\left(-x\right) \exp_{\kappa}\left(x\right) = 1\,,    
\end{equation}
\begin{equation}\label{symmetry}
    \exp_{-\kappa}\left(x\right) = \exp_{\kappa}\left(x\right)\,.
\end{equation}
The first one reflects the fact that these functions are inverses of each other, while the second one allows us to construct a generalized statistical. Indeed, the $\kappa$-generalized distribution is defined as~\cite{Kaniadakis_2013}
\begin{equation}
    f_i\sim\exp_{\kappa}\left(-\beta \varepsilon_i+\beta\mu\right),
\end{equation}
whose asymptotic limits give
\begin{equation}
    f_i\sim\left\{\begin{array}{cc}
        \exp\left(-\beta \varepsilon_i+\beta\mu\right),\ &\beta \varepsilon_i-\beta\mu \rightarrow0;\\
        \mathcal{N}\varepsilon_i^{-\frac{1}{k}},\ &\varepsilon_i\rightarrow\infty.\\
    \end{array}
    \right.
\end{equation}
Notice that it reproduces the Boltzmann ordinary distribution at low energy limit, whereas a power law tail is achieved to describe high energy systems. Finally, the last one tell us that the generalized exponential is symmetric with respect to the parameter $\kappa$. This means that symmetric values of $\kappa$ (e.g., $\kappa = \pm \, 0.5$) give rise to the same solution.

\begin{figure}
    \centering
    \includegraphics[scale=0.5]{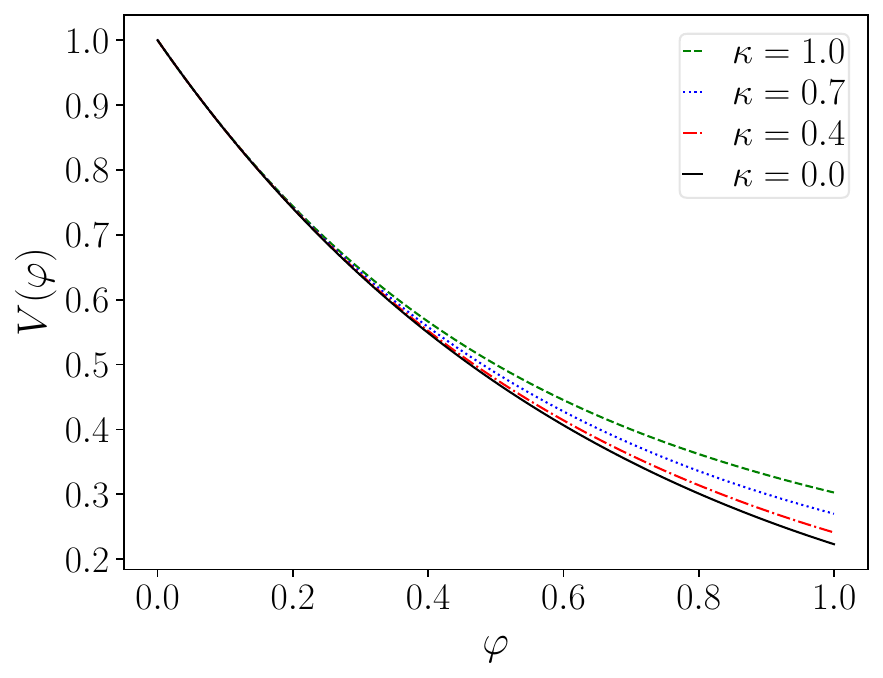}
    \caption{The generalized potential, eq.~(\ref{kpot}), as a function of the field $\varphi$ for some selected values of the deformation parameter $\kappa$ and $\lambda = 1.5$.}
    \label{kpotencial}
\end{figure}

We display in figure~\ref{kpotencial} the behavior of the generalized potential, eq.~(\ref{kpot}), as a function of the field $\varphi$. In particular, it must be noted that for all $\kappa \neq 0$ the generalized potential presents a quasi-exponential (power law) behavior, whereas for $\kappa=0$ the usual potential, eq.~(\ref{inflacaoexp}), is fully recovered. We also emphasize that because of the symmetry property highlighted in eq.~(\ref{symmetry}), we have considered only some selected positive values of $\kappa$ on this plot. However, in what follows, we will see that the straight relations between some inflationary parameters will result in functions that are not symmetric in $\kappa$. In these cases, we should also consider negative values of the parameter $\kappa$.

\subsection{Slow-roll results}
By considering the $\kappa$-generalized potential in eq.~(\ref{kpot}), the slow-roll parameters, eqs.~(\ref{epv})--(\ref{xiv}) give  
\begin{eqnarray}
    \epsilon &=& \frac{\lambda^2}{2} \frac{1}{[1 + \kappa^2 \lambda^2 \varphi^2]}\,, \label{HSSPV1} \\
    \eta &=& \frac{\lambda^2}{2}\frac{1}{[1 + \kappa^2 \lambda^2 \varphi^2]} \left(1 + \frac{2\kappa^2\lambda\varphi}{\sqrt{1 + \kappa^2 \lambda^2 \varphi^2}}\right)\,, \label{HSSPV2} \\
    \xi^2 &=& \frac{\lambda^4}{[1 + \kappa^2 \lambda^2 \varphi^2]^2}\left[(1 - \kappa^2) + \frac{3\kappa^2\lambda\varphi}{\sqrt{1 + \kappa^2 \lambda^2 \varphi^2}}\left(1 + \frac{\kappa^4\lambda\varphi}{\sqrt{1 + \kappa^2 \lambda^2 \varphi^2}}\right)\right]\,. \label{HSSPV3}
\end{eqnarray}
Hence, the number of {\it e}-folds defined in eq.~(\ref{e-folds}) gives
\begin{equation}\label{e-folds2}
    N = \frac{1}{2\lambda}\left[\varphi \sqrt{1 + \kappa^2 \lambda^2 \varphi^2} + \frac{1}{\kappa\lambda}\ln{(\sqrt{1 + \kappa^2 \lambda^2 \varphi^2} + \kappa\lambda\varphi)}\right]\,.
\end{equation}
As expected, the above expressions reduce to the usual exponential results $\epsilon = \eta = \lambda^2/2$, $\xi^2 =\lambda^4$, and $N = \varphi/\lambda$ in the limit $\kappa \rightarrow 0$.

\begin{figure}
\centering
\begin{tabular}{@{}c@{}}
\subfloat{\label{ep_h1}\includegraphics[width=0.475\linewidth]{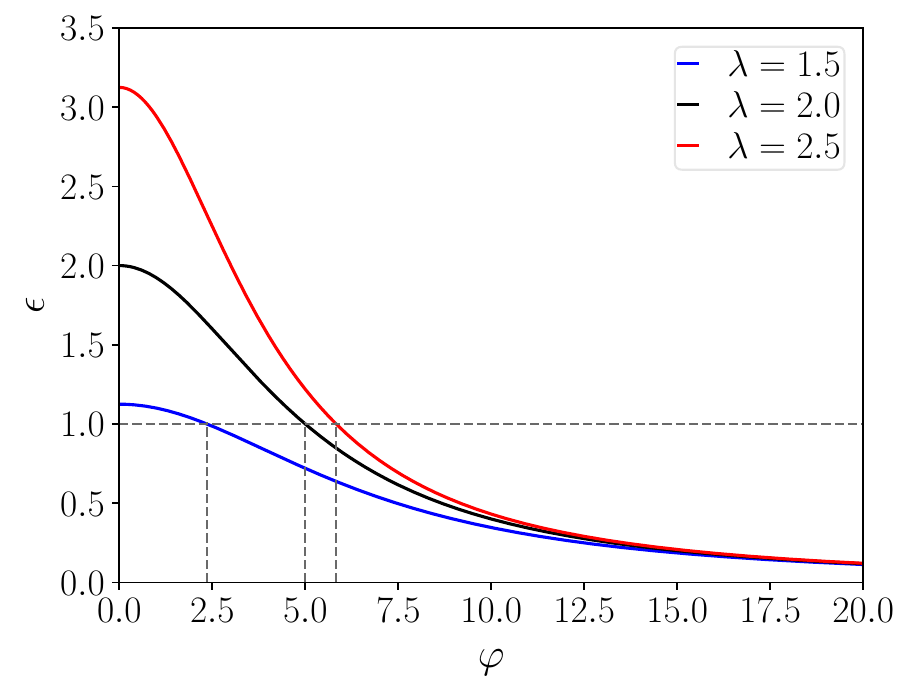}}\\ (a)
\end{tabular}\qquad 
\begin{tabular}{@{}c@{}}
\subfloat{\label{ep_h2}\includegraphics[width=0.475\linewidth]{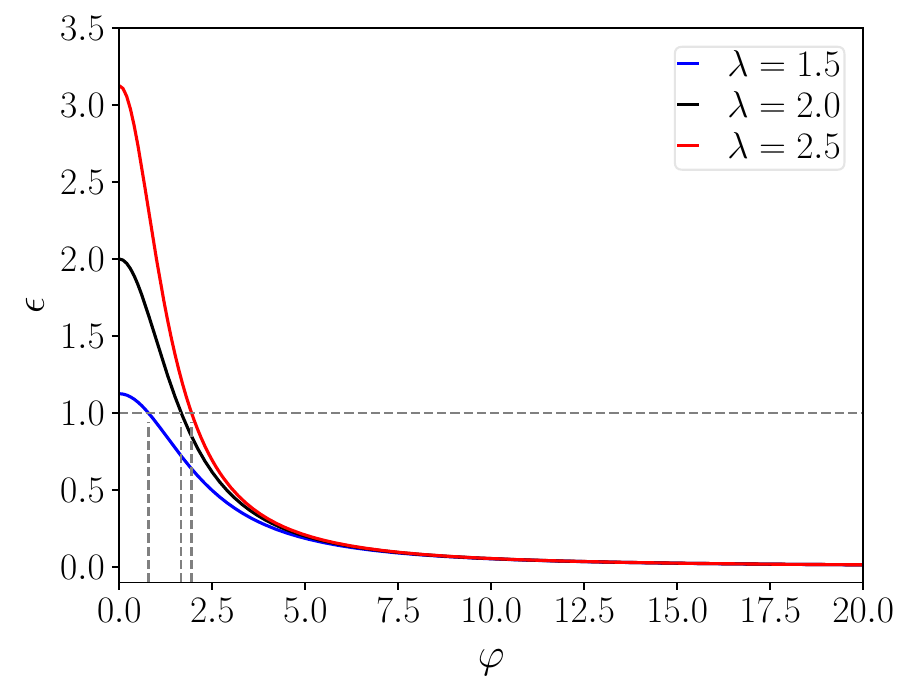}}\\ (b)
\\
\end{tabular}
\caption{The first slow-roll parameter $\epsilon$ as a function of the field $\varphi$ [eq.~(\ref{HSSPV2})] for selected values of $\lambda$ and (a) $\kappa=0.1$ and (b) $\kappa=0.3$. We see that increasing $\lambda$ also increases the value of the field at the end of inflation, $\varphi_\text{end}$, characterized by $\epsilon(\varphi_\text{end})=1$.}
\end{figure}

In addition, we can compute the value of the field at the end of inflation by setting $\epsilon (\varphi_\text{end}) = 1$, i.e., 
\begin{equation}\label{endinflation}
    \varphi_\text{end}=\frac{1}{\kappa}\left(\frac{1}{2}-\frac{1}{\lambda^2}\right)^\frac{1}{2},\quad \forall\ \kappa \neq 0 \, .
\end{equation}
For this value to be physically acceptable, i.e., $\varphi_\text{end} \in \mathbb{R}$, we must have $|\lambda| \geq \sqrt{2}$~, where we can disregard negative values of $\lambda$ as we are interested in only decreasing potentials. In this sense, we show in figure~\ref{ep_h1} the behavior of the slow-roll parameter $\epsilon$ as a function of the field $\varphi$ for selected values of $\lambda$ and $\kappa=0.1$; figure~\ref{ep_h2} is the same analysis, but for  $\kappa=0.3$. Notice that increasing $\lambda$ also increases the value of the field at the end of inflation in both cases, as we can see from eq.~(\ref{endinflation}).

Let us now consider both the inflationary parameters, spectral index and tensor-to-scalar ratio. Replacing the above results for the slow-roll parameters into eq.~(\ref{nsh}) and eq.~(\ref{ratio2}), we obtain
\begin{equation}\label{nsphi}
    n_{\rm s} - 1 = - \frac{\lambda^2}{[1 + \kappa^2 \lambda^2 \varphi^2]}\left(1 - \frac{2\kappa^2\lambda\varphi}{\sqrt{1 + \kappa^2 \lambda^2 \varphi^2}}\right)\,,
\end{equation}
\begin{equation}\label{rxk}
    r = 8\lambda^2 \frac{1}{[1 + \kappa^2 \lambda^2 \varphi^2]}\, ,
\end{equation}
and the straight relation between $n_{\rm s}$ and $r$,
\begin{equation}\label{planonsxr2}
    n_{\rm s} - 1=-\frac{r}{8}\left[1-2\kappa\left(1-\frac{r}{8\lambda^2}\right)^{\frac{1}{2}}\right]\,.
\end{equation}
Notice that the above eq.~(\ref{planonsxr2}) is the straight generalization of the well-known expression $r = 8\left(1 - n_{\rm s}\right)$, expected in the usual exponential scenario, as well as a scale invariant spectrum suggest vanishing primordial tensor modes. On the other side, the running of the spectral index gives
\begin{equation}\label{alpha2}
    \alpha_{\rm s} = 2 \lambda^4 \frac{\kappa^2}{[1 + \kappa^2 \lambda^2 \varphi^2]^2} \left[1 + \frac{\lambda\varphi}{\sqrt{1 + \kappa^2 \lambda^2 \varphi^2}}\left(1 - \frac{3\kappa^2\lambda\varphi}{\sqrt{1 + \kappa^2 \lambda^2 \varphi^2}}\right)  \right]\,,
\end{equation}
or, equivalently,
\begin{equation}\label{rxalphas}
    \alpha_{\rm s} = 2\kappa^2\left(\frac{r}{8}\right)^2 \left\{1 + \frac{1}{\kappa}\left(1 - \frac{r}{2\lambda^2}\right)^{\frac{1}{2}}\left[1 - 3\kappa\left(1 - \frac{r}{8\lambda^2}\right)\right]  \right\}\,,
\end{equation}
where we have replaced eq.~(\ref{rxk}) into eq.~(\ref{alpha2}) in order to obtain eq.~(\ref{rxalphas}). Notice that $\kappa=0$ implies $\alpha_{\rm s} = 0$. This means that the deformation parameter $\kappa$ makes the primordial fluctuations acquire a very small dependence on the scale according to $\alpha_{\rm s} \propto \kappa^2$.
\begin{figure}
\centering
\begin{tabular}{@{}c@{}}
\subfloat{\label{nsrplane}\includegraphics[width=0.475\linewidth]{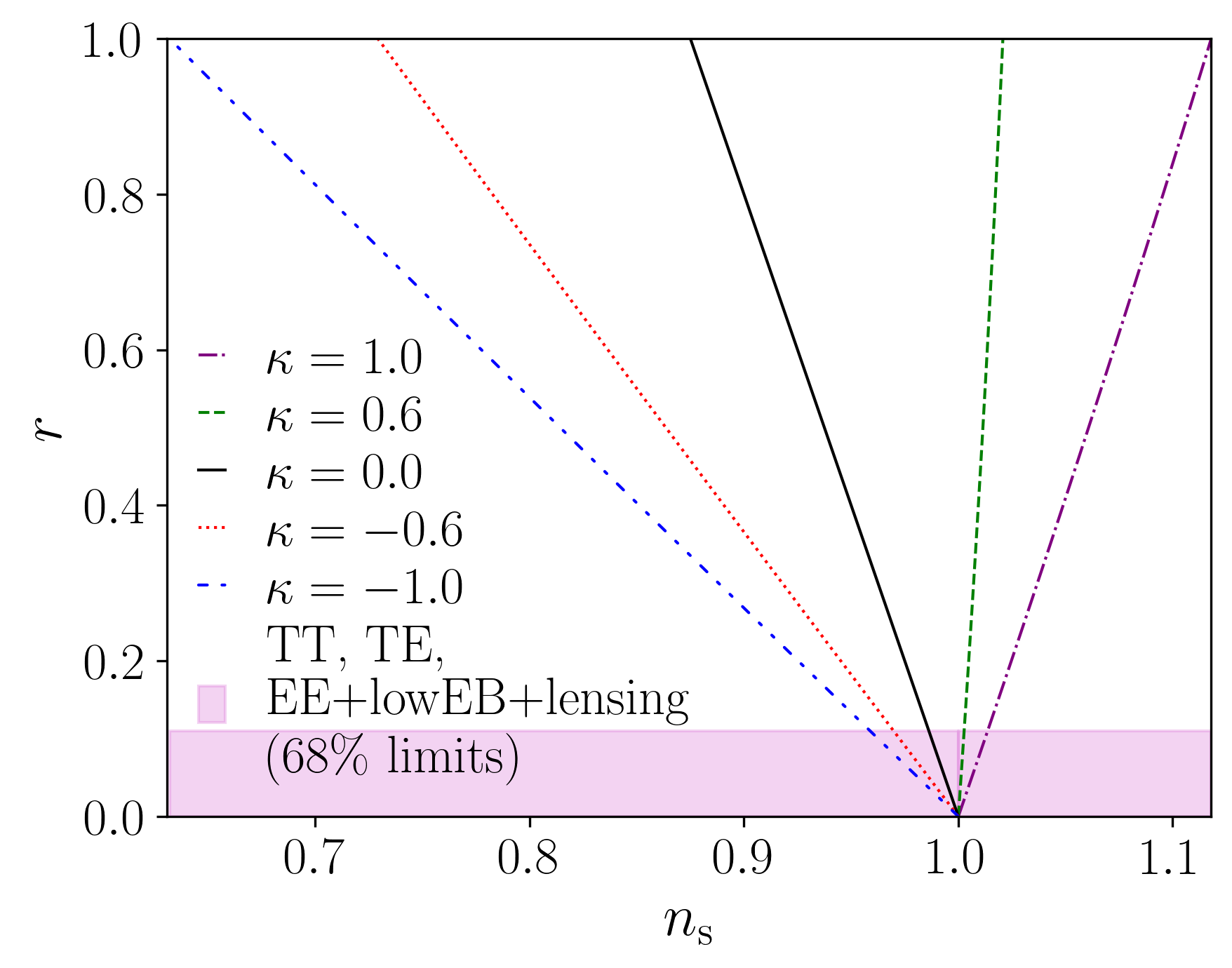}}\\ (a)
\end{tabular}\qquad 
\begin{tabular}{@{}c@{}}
\subfloat{\label{asnsplane}\includegraphics[width=0.475\linewidth]{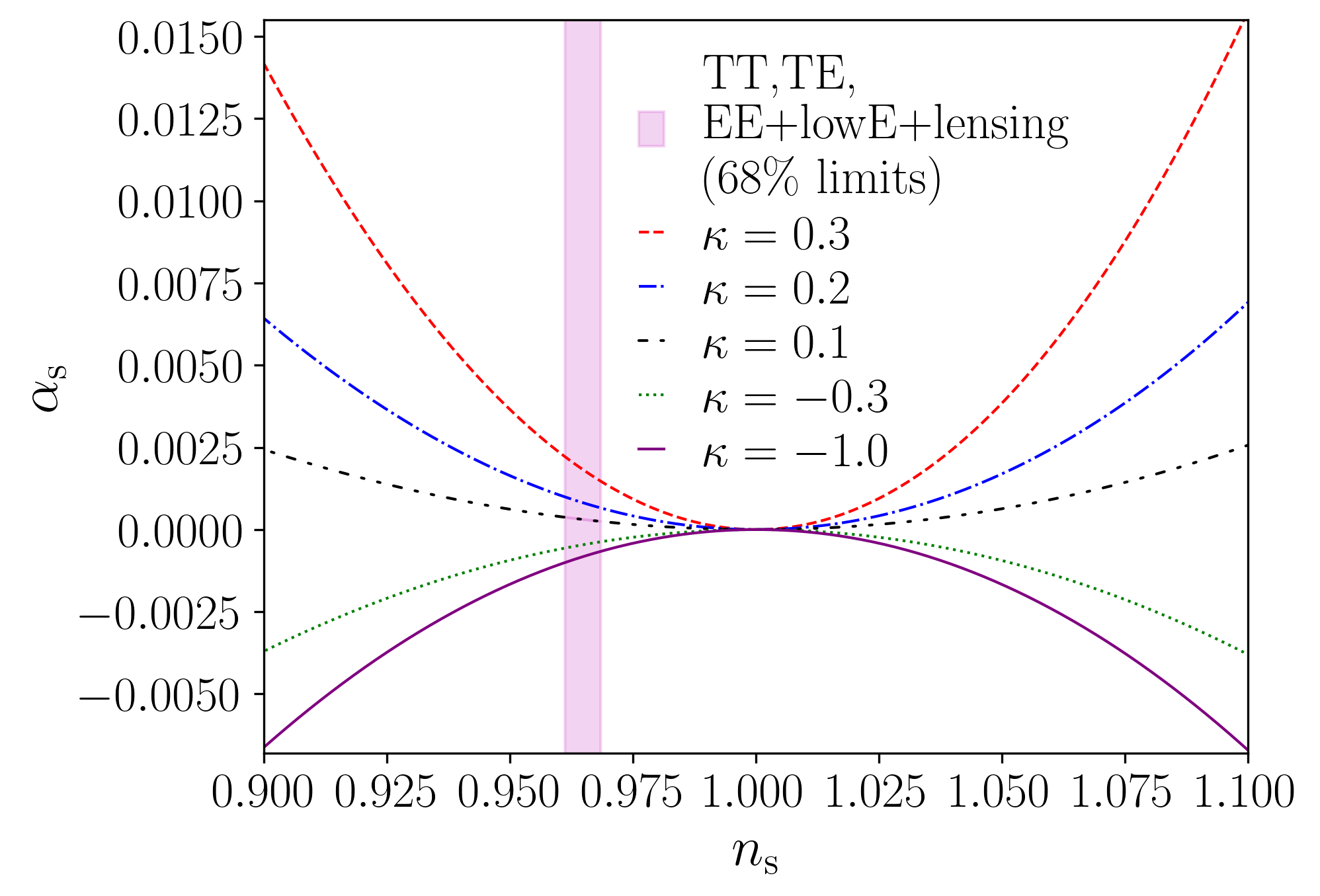}}\\ (b)
\\
\end{tabular}
\caption{Trajectories for different values of $\kappa$ on the planes (a) $n_{\rm s}-r$ and (b) $n_{\rm s} - \alpha_{\rm s}$. In (a) we consider the first order in the slow-roll approximation, eq.~(\ref{planonsxr2}); whereas (b) is the numerical solution envolving eqs.~(\ref{nsphi}) and~(\ref{alpha2}). In both cases, the hatched area corresponds to the current Planck plus lensing observations at $68\,\%$ CL \cite{Aghanim_2020}.}
\end{figure}
Next, we show in figures~\ref{nsrplane} and~\ref{asnsplane} the curves on the plane $n_{\rm s} - r$, originating from eq.~(\ref{planonsxr2}), and on the plane $n_{\rm s} - \alpha_{\rm s}$, corresponding to the numerical solution of eqs.~(\ref{planonsxr2}) and~(\ref{rxalphas}), respectively, for different values of the parameter $\kappa$ and $\lambda=1.5$. As we can see, for a sizable range of $\kappa$, the model's predictions in both cases are in fully agreement with current bounds from CMB plus lensing Planck data, $r < 0.11$ and $n_{\rm s} = 0.9649 \pm 0.0042$, at $68\,\%$ CL~\cite{Akrami_2020}. Notice that this time we also have used negative values of $\kappa$ since they provide different solutions than those considering positive $\kappa$. Notice also that, regardless of the value of $\kappa$, $n_{\rm s} = 1$ implies $r=0$ , which in turn implies $\alpha_{\rm s} = 0$.

\begin{figure}
\centering
\begin{tabular}{@{}c@{}}
\subfloat{\label{nsxk}\includegraphics[width=0.475\linewidth]{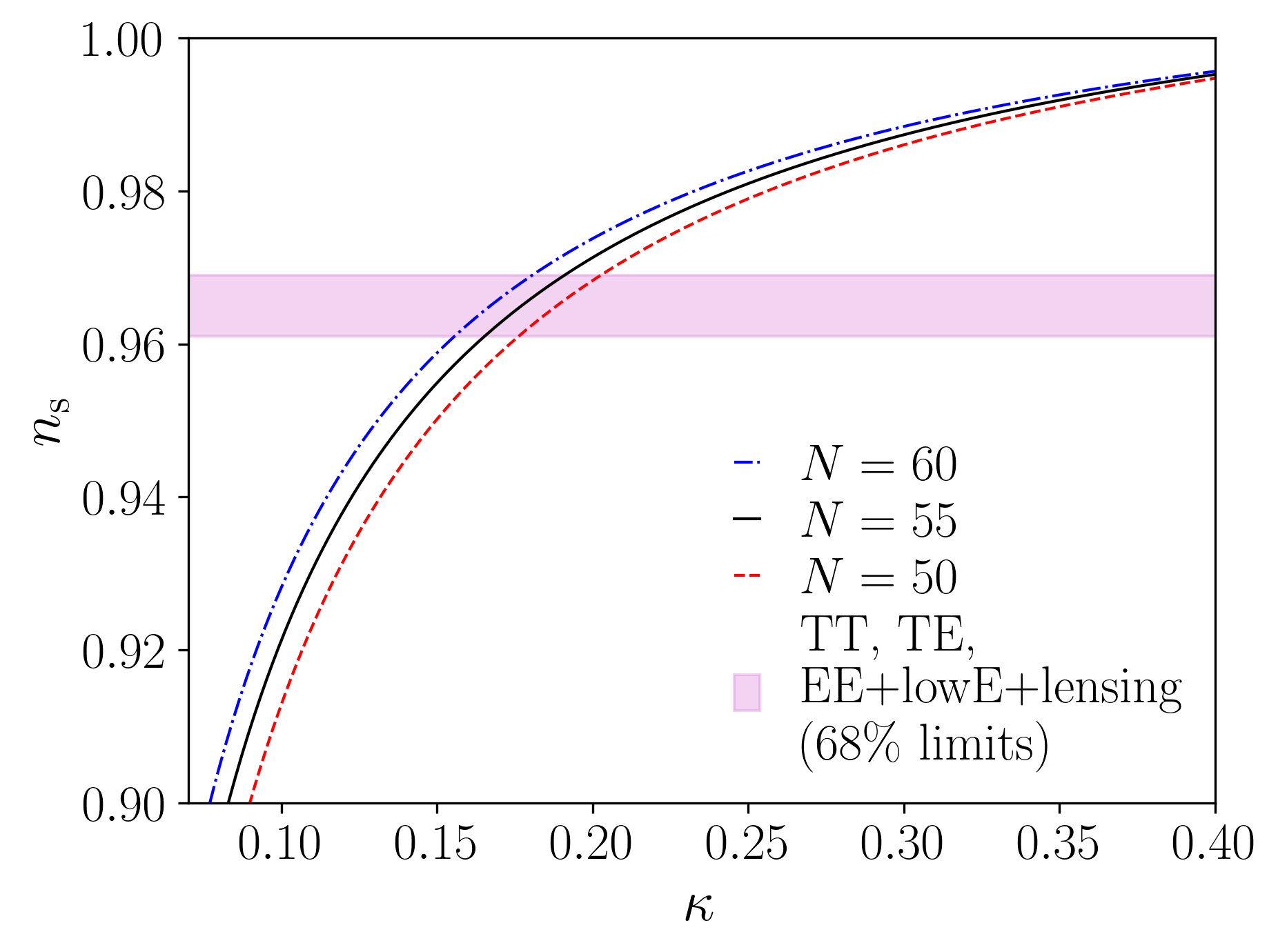}}\\ (a)
\end{tabular}\qquad 
\begin{tabular}{@{}c@{}}
\subfloat{\label{fNLxk}\includegraphics[width=0.475\linewidth]{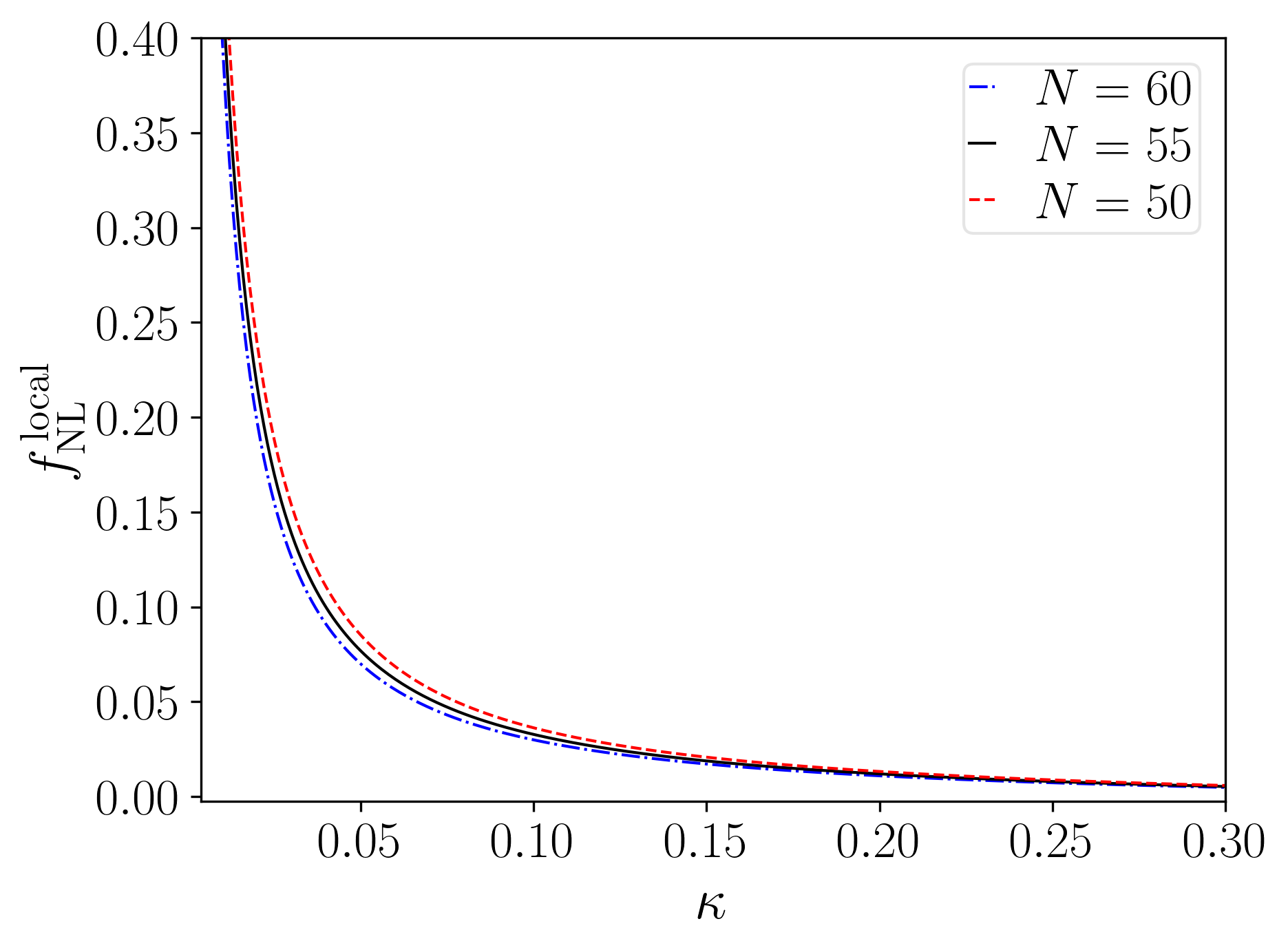}}\\ (b)
\\
\end{tabular}
\caption{(a) Spectral index $n_{\rm s}$ and (b) local non-linearity parameter $f_\text{NL}^\text{local}$ as functions of the parameter $\kappa$ for selected values of the number of {\it e}-folds ranging the interval $N = 55 \pm 5$ and $\lambda = 1.5$.}
\end{figure}

By combining eqs.~(\ref{e-folds2}) and~(\ref{nsphi}), we obtain the spectral index $n_{\rm s}$ as a function of $N$, so that we can compare the model's predictions with the current observation bounds on these quantities. It is important to point out again that both the PGWs and structure formation observations constraint $N$ to $50 - 60$ {\it e}-folds between the horizon exit and the end of inflation~\cite{Akrami_2020_3}. In figure~\ref{nsxk} we then show the solutions $n_{\rm s}(\kappa, N)$ for $N = \{50, 55, 60\}$. As we can see, only a small range of $\kappa > 0$ is in fully agreement with the Planck observations of $n_{\rm s}$ (hatched area).

Lastly, replacing eqs.~(\ref{HSSPV1}) and (\ref{HSSPV2}) into eq.~(\ref{fNLH}) we get the generalized non-linearity parameter,
\begin{equation}
    f_\text{NL}^\text{local} = \frac{5\lambda^2}{12} \frac{1}{[1 + \kappa^2 \lambda^2 \varphi^2]} \left(1 - \frac{2\kappa^2\lambda\varphi}{\sqrt{1 + \kappa^2 \lambda^2 \varphi^2}}\right).
\end{equation}
First, note the maximum value for the non-linearity parameter, $f_\text{NL}^\text{local} = 5\lambda^2/12$, corresponding to the usual exponential scenario $\kappa = 0$. Then, in a similar way to $n_{\rm s}(\kappa, N)$ -- or equivalently using the second equality in eq.~(\ref{fNLH}) --, the solutions $f_\text{NL}^\text{local}(\kappa, N)$ for $N = \{50, 55, 60\}$ are displayed in figure~\ref{fNLxk}. Since our $\kappa$-generalization has introduced the lower limit $\lambda \geq \sqrt{2}$, we can get large NGs (i.e., $f_\text{NL}^\text{local} \sim 1$) from very small values of $\kappa$ (i.e., $\kappa \lesssim 0.02$). This means that if the single field slow-roll approximation, characterized by the consistence relation in eq.~(\ref{fNLH}), is true, then we must have a $\kappa$ not so close to zero.

\section{\label{dic.}Discussion}
The inflationary paradigm represents one of the greatest successes of the fusion between the concepts of the quantum field theory and modern cosmology, bringing changes in the way we understand the primordial universe. It is the favorite hypothesis of most cosmologists to explain the origin of primordial fluctuations, being the basis for the emergence of structures, as well as for the flatness and statistical  isotropy of the universe. In this way, a number of inflationary models whose the physical motivations are rooted in the modern particle physics have been proposed.

In this paper we have investigated the main cosmological consequences of a new inflationary scenario characterized by the generalized potential in eq.~(\ref{kpot}).
This model is the result of the improvement of the exponential models, which reduce to the form of eq.~(\ref{inflacaoexp}), in the context of the $\kappa$-deformed theories~\cite{Kaniadakis_2001,Kaniadakis_2002,Kaniadakis_2013}.
As discussed in section~\ref{k-exp.inf.}, this generalized potential behaves like a simple power law for all $\kappa \neq 0$ and exactly like an exponential function for $\kappa = 0$, as we can see in figure~\ref{kpotencial}.

From eq.~(\ref{endinflation}) we have found that $|\lambda| \geq \sqrt{2}$, where we have disregarded negative values as we are only interested in decreasing potentials. As it happens, the scalar field $\varphi$ must be a real quantity after all. This constraint has straight influence on the energy scale of inflation, as well as on the slow-roll parameter $\epsilon$, which signals the end of inflation. Indeed, increasing $\lambda$ also increases the value of the field at the end of inflation, as we can see in eq.~({\ref{endinflation}}) and figures~\ref{ep_h1} and \ref{ep_h2}.

Next, we have analyzed the scalar spectral index, its running, and tensor-to-scalar ratio in light of the $n_{\rm s} - r$ and $n_{\rm s} - \alpha_{\rm s}$ planes. The straight relation between $n_{\rm s}$ and $r$ is given in eq.~(\ref{planonsxr2}), while for $n_{\rm s}$ and $\alpha_{\rm s}$ we have needed to numerically solve eqs.~(\ref{planonsxr2})--(\ref{rxalphas}). Figures~\ref{nsrplane} and \ref{asnsplane} show us the solutions $r(n_{\rm s})$ and $\alpha_{\rm s}(n_{\rm s})$ for different values of the parameter $\kappa$ and $\lambda = 1.5$, respectively. In both cases, the hatched area corresponds to the current Planck plus lensing observations at $68\,\%$ CL. As we can see, the model's prediction for these parameters are compatible with CMB and LSS observational constraints.

We also have analyzed the spectral index $n_{\rm s}$ as a function of the parameter $\kappa$ for three different values of the number of {\it e}-folds corresponding to the range $N = 55 \pm 5$ and $\lambda = 1.5$. Figure~\ref{nsxk} shows us that only small range of $\kappa$, including only positive values, is compatible with the current bounds on $n_{\rm s}$ from Planck plus lensing data, $n_{\rm s} = 0.9649 \pm 0.0042$ at $68\,\%$ CL (hatched area).

Since the single field slow-roll inflationary mechanism is expected to produce small (but detectable) primordial NGs, we also have investigated possible non-vanishing contributions to the three-point correlation function encoded in the local non-linearity parameter. In figure~\ref{fNLxk} we show the behavior of $f^\text{local}_\text{NL}$ as a function of the parameter $\kappa$ for $N = \{50, 55, 60\}$ and $\lambda = 1.5$. As we can see, $f^\text{local}_\text{NL}$ has a maximum value when $\kappa = 0$ (usual scenario) and a minimum for $\kappa \rightarrow 1$. The maximum one depends only on $\lambda$, which obeys $\lambda \geq \sqrt{2}$. Hence, we can get large local NGs, i.e., $f^\text{local}_\text{NL} \gtrsim 1$, for very small $\kappa$, tipically $\kappa \lesssim 10^{-2}$. This means that, if the single field slow-roll inflation is the true mechanism for generating the primordial fluctuations, we expect a $\kappa$ not so close to zero.

However, there is a tenacious observational nuisance when it comes to primordial NGs, since not even the most current and accurate measurements of $f_\text{NL}$ (be it local, equilateral or orthogonal) allow us to make any statements about it. It turns out that there are many sources of NGs in the CMB anisotropies beyond the primordial one, including systematics effects and astrophysical contamination~\cite{Bartolo_2012}. Well-known examples of contamination are those produced by secondary non-linear anisotropies, generated since the epoch of matter-radiation decoupling at $z \sim 1100$, namely the integrated Sachs-Wolfe effect, gravitational lensing, Sunyaev-Zel’dovich effect, polarization, among others~\cite{Aghanim_2008}. In addition, another important source of NG in the CMB anisotopies are the non-linearities in the evolution of the photon-baryon fluid at recombination~\cite{Bartolo_2009}.

As a conclusion, we highlight the $\kappa$-exponential inflation as a promising model describing the origin of the primordial fluctuations, whose the main predictions such as the number of {\it e}-folds, tensor-to-scalar ratio, scalar spectral index and its running, and primordial NGs are in fully agreement with the most current and accurate Planck measurements. Our theoretical results tell us that the value of $\kappa$ compatible with the observations depends on the analysis, e.g., $n_{\rm s}-r$ plane analysis points out to $\kappa < 0$, $n_{\rm s}-\kappa$ to $\kappa >0$, and $\alpha_{\rm s}-n_{\rm s}$ to both. Meanwhile, the $f_\text{NL}^\text{local}-\kappa$ analysis discards negatives and very close to zero values of $\kappa$. Finally, combining the $n_{\rm s}-\kappa$ and $f_\text{NL}^\text{local}-\kappa$ analyses we conclude that $0.02 \lesssim \kappa \lesssim 0.2$. It is worth remembering, in the analyzes involving both $n_{\rm s}-k$ and $f_\text{NL}^\text{local}-k$ planes we have considered three values of the number of {\it e}-folds, $N = \{50, 55, 60\}$.


\acknowledgments

The authors are grateful to the Brazilian agency CAPES for financial support and to the CENAPAD-SP for computational support.


\end{document}